\documentstyle[12pt,aaspp4]{article}

\def\Msun{M$_{\odot}$}
\def\Lsun{L$_{\odot}$}

\begin{document}

\paperid{MS-96-5172}
\received{12 March 1996}
\accepted{14 June 1996}

\title{Limits on the Halo White Dwarf Component of Baryonic Dark
Matter from the {\em Hubble Deep Field}}

\author{Steven D. Kawaler \altaffilmark{1}}
\altaffiltext{1}{Department of Physics and Astronomy, Iowa State University, 
Ames, IA  50011  USA; sdk@iastate.edu}

\begin{abstract}

The MACHO collaboration lensing event statistics suggest that a significant
fraction of the dark galactic halo can be comprised of baryonic matter in the
form of white dwarf stars with masses between 0.1 and 1.0 \Msun .  Such a
halo white dwarf population, in order to have escaped detection by those who
observe the white dwarf luminosity function of the disk, must have formed
from an old population.  The observations indicate that the number of halo
white dwarfs per cubic parsec per unit bolometric magnitude is less than
$10^{-5}$ at $10^{-4.5}$\Lsun; the number must rise significantly at lower
luminosities to provide the needed baryonic halo mass.  Such white dwarfs may
easily escape detection in most current and earlier surveys.  Though it is
limited in angular extent, the {\em Hubble Deep Field} (HDF) probes a
sufficient volume of the galactic halo to provide interesting limits on the
number of halo white dwarf stars, and on the fraction of the halo mass that
they can make up.  If the HDF field can be probed for stars down to $V=29.8$
then the MACHO result suggests that there could be up to 12 faint halo white
dwarfs visible in the HDF.  Finding (or not finding) these stars in turn
places interesting constraints on star formation immediately following the
formation of the galaxy.

\end{abstract}


\section{Introduction}

Two recent investigations that have exciting consequences over a broad
spectrum of modern astrophysics were announced at the American Astronomical
Society meeting in January 1996.  First, the MACHO collaboration released a
summary of their observations of microlensing events in the Milky Way,
interpreting the profiles of 7 long-duration lensing events in the direction
of the LMC as evidence for lensing by objects with masses between 0.1 and 1.0
\Msun (Bennett et al.  1995, 1996).  The lensing objects must be of extremely
low luminosity; therefore ordinary dwarf stars are ruled out.  They suggest
that about 50\% of the dark matter halo of the Milky Way could therefore be
comprised of white dwarf stars.  Considering the large mass of the galactic
halo, this implies that halo white dwarfs must be extremely abundant.  Direct
observational constraints on the halo white dwarf luminosity function from
observations (Liebert et al.  1988) are not necessarily in conflict with the
MACHO result; however the luminosity function of white dwarfs in the disk
must then contain a fair fraction of halo white dwarfs.

Given the age of the halo as determined from, for example, globular cluster
studies, the observed downturn in the white dwarf luminosity function at
$\log(L/L_{\odot})\approx-4.5$ ($M_{\rm bol}\approx 16.0$, with no white
dwarfs with $M_{\rm bol} > 16.2$) constrains the star formation history for
the generation of halo stars that might have produced such a halo white dwarf
population (Tamanaha et al 1990, Adams \& Laughlin 1996).  Under conventional
assumptions about star formation early in the history of the galaxy, Adams \&
Laughlin (1996) conclude that the observations of Liebert et al. (1988)
already limit the fraction of the dark matter halo that can be attributed to
white dwarfs to less than 25\%.  On the other hand, Tamanaha et al. (1990)
show that extreme conditions, such as an enormous burst of star formation
early in the history of the galaxy, could produce a massive number of white
dwarfs in the halo that could escape detection.

The second investigation summarized at the meeting was the Hubble Deep Field
(HDF) (Williams et al. 1995).  The HDF is an extremely deep set of images
obtained with the Hubble Space Telescope in a relatively blank field out of
the galactic plane.  The HDF has quoted $3\sigma$ detection limits of
approximately 30th magnitude in the $V$ and $I$ plates.  Though a very narrow
angle survey, the depth of the HDF results in its sampling a significant
volume of the halo.  Thus it is useful for the purposes of detecting (or
placing upper limits on the distribution of) halo white dwarf stars.  This
Letter discusses the use of the HDF to constrain the luminosity function of
halo white dwarfs.  It outlines the computation of the number of white dwarfs
expected in the HDF as a function of the limiting magnitude for detection of
stars and the luminosity of the faintest expected white dwarfs.  As discussed
in the final section, identification of stellar candidates to near the limit
of sensitivity of the HDF will provide significant constraints on the
fraction of the halo mass that can be attributed to white dwarf stars, or on
the luminosity function of such stars.

\section{The Halo as Probed by the HDF}

For a small survey area such as the HDF, the volume $V$ of space sampled 
out to a distance $d$ (in pc) given an area of $A$ square arc minutes is 
simply
\begin{equation}
 V = 2.82\times 10^{-8} Ad^3 \; \; {\rm pc}^3 \; .
\end{equation}
The HDF covered an area of approximately 4 square arc minutes; therefore 
since it looked out of the plane, taking $d=500$~pc suggests that it 
samples a disk volume of only 14~pc$^3$.  With such a small volume sample, 
it is not surprising that no white dwarfs are in the field; the approximate 
space density of white dwarf stars is $3\times 10^{-3}$ per cubic parsec 
(Liebert et al.  1988).

For a putative population of white dwarfs in the halo, the HDF has sampled a
much larger volume.  Assuming a limiting magnitude of $m_l$, and a white
dwarf with an absolute magnitude $M_f$ (in the same band as the limiting
magnitude), then the distance out to which that white dwarf would be visible,
$d_f$, is simply
\begin{equation}
\log(d_f)=0.2 (m_l-M_f+5)\;\;.
\end{equation}
With this value for the range of the survey, a lower limit for the effective
volume contained within can be written in terms of the limiting magnitude of
the survey and the absolute magnitude of the faintest white dwarf:
\begin{equation}
\log(V)= -4.550 + \log(A) + 0.6(m_l-M_f)
\end{equation}
where $V$ is in pc$^3$ and $A$ is in square arc minutes.

Another way that the depth of a survey can be judged is by assuming a mass 
distribution for the halo, and then computing the halo mass contained 
within the survey volume.  Thus if the density distribution is assumed to 
be spherically symmetric about the center of the galaxy, (i.e. $\rho=\rho_o
f(r)$) then the mass (in \Msun) contained within the survey area is
\begin{equation}
M_H = 9.4\times 10^{-9} A \rho_o \int_{0}^{d_f} f(r)\delta^2 d\delta
\end{equation}
The distance from the galactic center $r$ can be expressed in terms of
the integration variable $\delta$, the distance from the Sun, with the
following transformation:
\begin{equation}
 r^2 = \delta^2 + D^2 -2 \delta D \cos b \cos l 
\end{equation}
where $(b,l)$ is the direction of the survey in galactic coordinates, and $D$
is the distance between the Sun and the galactic center.

For the halo structure function $f(r)$, we adopt the standard form used by 
Binney \& Tremaine (1987):
\begin{equation}
f(r)=\frac{1}{1+(r/a)^\gamma}
\end{equation}
where $a$ and $\gamma$, along with $\rho_o$, are parameters derived from
dynamical studies of the Galaxy.  For simplicity we take $\gamma=2$ which
allows analytic integration for the mass.   With this form for the halo mass
distribution, the mass of the halo sampled by the HDF field (in solar
masses) can be written in terms of $d_f$ after a bit of integration 
(and a lot of algebra) as
\begin{eqnarray}
M_H & = &  16.07 A \rho_o a^2 \nonumber \\
 & \times & \left\{ d_f  -   \frac{B}{2}\ln \left(1+\frac{d_f^2}{E^2}
                                    +\frac{Bd_f}{E^2} \right) \right. \\
 &   & + \frac{B^2-2E^2}{C}\left[\arctan \left\{\frac{2d_f+b}{C}\right\}
       \left.    -\arctan \left\{\frac{B}{C}\right\} \right] \right\} \nonumber
\end{eqnarray}
where
\begin{eqnarray}
E^2 & = &a^2+D^2 , \\
B & = & -2D \cos b \cos l , 
\end{eqnarray}
and
\begin{equation}
C = \sqrt{4E^2-B^2}  .
\end{equation}

The values of $B$, $C$, $D$, $E$, $a$, and $d_f$ are all in parsecs in the
above expressions.  For the Hubble Deep Field, $b \approx 54^o$ and $l
\approx 127^o$.  This fixes $B$ as 6.01~kpc.  The values of $a$ and $\rho_o$
depend on the precise model favored for the halo, but representative values
are $a=2$~kpc and $\rho_o=0.19$~\Msun/pc$^3$, from Bahcall \& Soneira
(1980).  Alternate halo models which preserve the $M/L$ ratio in the disk and
the total halo mass out to $r=D$ are possible.  For the representative values
of the above parameters, the value of $M_H$ rises from 1.05\Msun~for
$d_f$=1.0~kpc to 7.8\Msun for $d_f$=2.0~kpc; Table~1 shows $M_H$ as a
function of $m_l-M_f$.  Thus the HDF in principle samples several solar
masses of halo material.

\section{Halo White Dwarfs and the HDF}

If, as suggested by the MACHO results, the halo dark matter of the Milky Way
is up to 50\% (by mass) halo white dwarfs, then up to half of the halo mass
sampled in the HDF can be white dwarf stars.  The mass sampled is dependent
upon $d_f$ which in turn is set by the absolute magnitudes of the white
dwarfs and the limit of detectability in the images. The expected number of
white dwarfs will be a fraction $x_f$ of the total number that are more
luminous than the magnitude limits used to determine $d_f$.  The number of
white dwarfs that should be visible on the HDF can be estimated by assuming a
reasonable mean mass for white dwarfs; Table~1 gives these numbers for a mean
mass of 0.6\Msun\/ in terms of $x_f$, assuming that 50\% of the halo mass in
the field is attributable to white dwarfs.

The fraction of halo white dwarfs brighter than a given $M_{\rm bol}$, $x_f$,
contains the dependency of the results on the white dwarf luminosity function
and therefore on the age of the halo, the history of star formation there,
and the physics of white dwarf cooling (see, for example, Wood 1992).  Here
we consider a few simplified cases shows the range of possible values of
$x_f$.  In these cases, we simplify the inputs to the luminosity function by
considering a single--mass population of white dwarfs (with $M=0.6M_{\odot}$)
and ignore the (small, Adams \& Laughlin 1996) effects of a distribution of
initial masses, the pre-white dwarf lifetimes, etc.  Under this
approximation, the luminosity function is inversely proportional to the
fading rate $d(\log L)/dt$, and in the event that $t_{\rm cool} = K
L^{\alpha}$, $x_f$ is approximated by
\begin{equation}
x_f = \frac{K L_{\rm lim}^{\alpha}-t_{\rm young}}
           {t_{\rm old}-t_{\rm young}}\;, 
\end{equation}
where $L_{\rm lim}$ is the minimum luminosity white dwarf detectable, $t_{\rm
old}$ is the age of the oldest white dwarf in the halo, and $t_{\rm
young}$ is the age of the youngest white dwarf in the halo.  Though
crude, these simplifications provide a decent estimate of the range of
$x_f$.  Also, the bolometric correction is problematic for the coolest white
dwarfs (Liebert et al. 1988, Wood 1992) and so it is assumed to be zero here,
in line with the arguments presented in Liebert et al. (1988).

In the first limit, assume that white dwarfs cool via the Mestel (1952) law
($t_{\rm cool} = 5.5\times 10^6  L^{-5/7}$~yr) normalized to give a cooling time
for white dwarfs in the disk of $9\times 10^9$~yr (Hansen \& Kawaler 1994).
In this limit, halo white dwarfs formed 16~Gyr ago have faded to a luminosity
of $10^{-4.85}$\Lsun, or to approximately 0.9 magnitudes fainter than the
disk cutoff.  In this case, $x_f=1$ out to where the HDF can see white dwarfs
with $M_V=17.1$ or brighter.  From Table~1, if the limiting $V$ magnitude of
the HDF is 28.8, then $x_f=1$ out to 2.2~kpc.

A second (more realistic) case allows for more rapid fading of white dwarfs
at low luminosities (below $10^{-4.5}$ \Lsun) as the result of
crystallization and Debye cooling.  In this phase, a rough fit to the models
of Winget et al.  (1987), gives $t_{\rm cool} \approx 2.85\times 10^8
L^{-1/3}$~yr; halo white dwarfs with an age of 16~Gyr have faded
approximately 1.9 magnitudes below the disk cutoff.  Assuming that star
formation in the halo was continuous until the disk formed, 40\% of halo
white dwarfs have absolute magnitudes brighter than $M_V=17.1$, compared with
100\% in the previous limit.  An upper limit to $x_f$ in this case, with the
same limiting $V$ and $M_V$ as above, is $x_f=0.4$ at 2.2~kpc.  Of course, if
star formation in the halo declines significantly between the initial
collapse phases and the formation of the disk, the value of $x_f$ would be
further reduced.  As an example, if halo star formation began 16 Gyr ago and
lasted 2 Gyr, then the most luminous halo white dwarf would have $M_V\approx
17.6$, and (with a $V$ magnitude limit on the HDF of 28.8) $x_f$ would be
zero beyond 1.7~kpc ($m_l-M_f > 11.2$).  A younger halo age increases $x_f$
for a given $m_l-M_f$; if the halo is only 12~Gyr old and star formation
continued for 2~Gyr, the most luminous halo white dwarfs would then have
$L=10^{-4.64}$\Lsun, and $x_f$ would be 0.94 under the conditons above
($m_l=28.8, M_f=17.1$).

Other more extreme cases can be considered, such as rapid and very intense
bursts of star formation early in the history of the halo with a subsequent
exponential decline in the birth rate with a time scale of less than 1~Gyr
(Tamanaha et al. 1990). In such models, the luminosity function of halo white
dwarfs can increase rapidly at luminosities below the disk cutoff, and still
be consistent with the observations of Liebert et al. (1988).  Parameters of
these burst models include the age of the galaxy; if for example the age of
the galaxy is significantly larger than 14 Gyr, then $x_f$ would be
correspondingly small; the earlier the burst, the smaller $x_f$ is.

In all of the above cases, the value of $x_f$ depends on the age of the halo;
if the halo is 16~Gyr or older, then halo white dwarfs would have had
sufficient time to fade below detectability on the HDF.  However, since some
current cosmological investigations favor an age for the Universe that is
comparatively young (i.e. Freedman, et al. 1994), the possibility remains
that a significant number of halo white dwarf stars population can be
detected on the HDF if indeed they make up a significant fraction of the
baryonic component of the dark halo of our Galaxy.  Assuming then that there
are a significant fraction of white dwarfs with $M_{\rm bol}<18$, and
considering the $3\sigma$ limiting magnitude of 30 in the red bands of the
HDF, a total white dwarf population of $12 x_f$ objects might exist in the HDF.

One can now ask if such halo white dwarfs should have been seen in studies 
of the white dwarf luminosity function that were restricted to the solar 
neighborhood.  This issue has been addressed by Adams \& Laughlin (1996) 
who show that the luminosity function of white dwarfs in the solar 
neighborhood as reported by Liebert et al.  (1988) places strict upper 
limits on the numbers of halo white dwarfs at luminosities above the disk 
cutoff luminosity of $10^{-4.5}$\Lsun.  The last data points in the disk 
white dwarf luminosity function are upper limits of approximately $5\times 
10^{-5}$pc$^{-3}M_{\rm bol}^{-1}$ at this luminosity.  For luminosity functions 
that derive from standard models of cooling white dwarfs, Adams \& Laughlin 
(1996) show that their luminosity function must quickly rise to several 
$\times 10^{-4}$pc$^{-3}M_{\rm bol}^{-1}$ and remain high down to very low 
luminosities if the age of this population is not excessively larger than 
the age of the oldest globular clusters.

Is the space density of the halo white dwarfs sufficiently low that it is not
in conflict with the observed cutoff for disk stars?   Assuming that the dark
halo density is tracked by halo white dwarfs seen by MACHO, then the local
space density of visible halo white dwarfs (that is, at the solar distance
from the galactic center) should be given by $0.5 \rho_o f(8.5)
\rm{kpc})/<m_{\rm wd}>$.  This is a space density of $8.3 \times 10^{-3} x_f$
halo white dwarfs per cubic parsec in the solar neighborhood.  This places
the nearest halo white dwarf about 5 pc away, on average.

\section{Summary and Conclusions}

There are no obvious stars in the HDF images (apart from a few ``bright''
20th magnitude stars); the task of discriminating between stellar and
nonstellar objects at very faint magnitudes requires extreme care.  Still, if
halo white dwarfs are old and therefore quite faint and red, there could
indeed be halo white dwarfs hiding at slightly fainter magnitudes in the HDF
-- though not as many as suggested by the MACHO result (Adams \& Lauglin
1996).  A recent preprint by Flynn et al.  (1996) reports that no stellar
objects exist down to $V=26.3$ for objects with $2.5 > V-I > 1.8$.  For white
dwarf stars, $V-I$ at the cool end of the white dwarf luminosity function is
a the red end of this range.  With a conservative value for the absolute
magnitude of representative halo white dwarfs of $M_f=18.0$ (i.e. ages of
approximately 16 Gyr), this limit to $V$ of 28.8 corresponds to $m_l-M_f$ of
10.8.  Table~1 shows that approximately three halo white dwarfs should have
been seen {\em if $x_f=1$}.  From the discussion of the previous section,
$x_f$ can range from 1 to nearly zero, depending on the population of halo
stars that produced white dwarfs.  A value of $x_f$ of 0.4 is likely; thus
Flynn et al. (1996) should have seen at least one or two white dwarfs.  The
fact that they saw none, though it fails to independently confirm the MACHO
suggestion, is inconclusive.  The dependence on $x_f$ makes this a weak
constraint, though models of star formation in the halo can easily produce
more luminous white dwarfs than the $M_V=18$ used above with nonzero values
of $x_f$.

If the MACHO suggestion is correct, then a significant number of white dwarfs
(3 times as many) could be detected if one could reach a magnitude limit for
stellar objects that is fainter by one more magnitude, to $I=27.3$, or
$V=29.8$ on the V exposures.  Given the signal--to--noise statistics
associated with the HDF exposures, along with judicious morphology and color
cuts, such limits should be approachable.  If the HDF still fails to yield
white dwarf stars, then either 1) the MACHO suggestion that 50\% of the mass
of the baryonic halo is in the form of white dwarfs is incorrect, or 2) the
value of $x_f$ is significantly smaller than predicted with conventional
models of white dwarf formation and evolution in the halo.  Already,
observations of the disk white dwarf luminosity function require that $x_f$
be small enough to allow a large enough number of white dwarfs in the solar
neighborhood, as follows from the work of Adams \& Laughlin (1996).  Point 2
would in turn indicate that acceptance of the MACHO hypothesis implies a
large and fast burst of star formation in the early years of the evolution of
our Galaxy in excess of 16~Gyr ago, or that white dwarfs cool more quickly 
than conventional theoretical models predict below $M_{\rm bol} \approx 18.0$.

\acknowledgements

This work was supported in part by an NSF Young Investigator Award (Grant 
AST-9257049).  The author wishes to thank the referee for several helpful
comments that have improved the presentation of this subject, and Marshall
Luban for illuminating discussions.

\clearpage


\begin{table}
\begin{center}
\caption{Number of halo white dwarfs expected on HDF.}
\vspace*{0.5in}
\begin{tabular}{cccc}

$m_l-M_f$  & $d_f$ [kpc] & $M_H$ [\Msun] & $N_{wd}/x_f$ \\
\tableline

   9.50  &    0.794  &    0.536  &   0.446 \\    
   10.0  &    1.000  &    1.054  &   0.878   \\  
   10.5  &    1.260  &    2.065  &   1.720   \\ 
   11.0  &    1.585  &    4.022  &   3.352   \\ 
   11.5  &    1.995  &    7.780  &   6.483   \\ 
   12.0  &    2.512  &    14.91  &   12.42   \\ 
   12.5  &    3.162  &    28.21  &   23.51   \\ 
   13.0  &    3.981  &    52.54  &   43.78   \\ 
   13.5  &    5.012  &    95.93  &   79.95    
\end{tabular}
\end{center}
\end{table}

\end{document}